\documentstyle[sprocl,psfig]{article}

\def\be{\begin{equation}}
\def\ee{\end{equation}}
\def\bea{\begin{eqnarray}}
\def\eea{\end{eqnarray}}
\def\alt
 {{\tiny \begin{array}{l}  <\\ \sim \end{array}}} 
\def\agt
 {{\tiny \begin{array}{l}  >\\ \sim \end{array}}}

\begin{document}

\title{
IMPLICATIONS OF PLANCK AND MAP MEASUREMENTS
ON SPARTICLE SPECTRA
}

\author{R. ARNOWITT}

\address{Center for Theoretical Physics, Department of Physics,\\
Texas A\&M University, College Station, TX 77843-4242, USA}

\author{PRAN NATH}

\address
{Department of Physics, Northeastern University, Boston, MA 02115, USA}


\maketitle\abstracts{
Future sattelite (MAP and Planck), balloon and ground based experiments 
will determine the basic cosmological parameters within a few percent.
We examine here the effect of this on constraining the SUSY parameter space 
for supergravity $R$-parity conserving models (with $\tan\beta \le 25$) 
for the cases of  $\nu$CDM and $\Lambda$CDM cosmological models.
For the $\nu$CDM ($\Lambda$CDM) models, the gluino mass is restricted 
by $m_{\tilde{g}}\alt 720(540) GeV$.
In both cases, the cosmological constraints are sensitive to non-universal 
SUSY soft breaking producing a lower bound $m_{\tilde{g}}\agt 400 GeV$
in some regions and for the $\nu$CDM model, gaps in the allowed 
$m_{\tilde{g}}$ range for other regions.
For gluino (neutralino) masses greater than $450(65) GeV$, $m_0$ is 
constrained to be small making squark and slepton masses generally 
light and determined mostly by  $m_{\tilde{g}}$.
}

\section{Introduction}
\label{sec:1}

One of the significant phenomena that has arisen with the development 
of supersymmetry has been the deepening of the connection between 
particle physics and cosmology.
While this connection has always been present (e.g. from the early 
nuclear synthesis calculations \cite{1}) supersymmetry (SUSY) 
now offers models that reach up in energy to the GUT scale 
$M_G\approx 10^{16} GeV$ (and perhaps further to the Planck scale) 
and backwards in time to the very early universe. 
In particular, models with $R$-parity invariance automatically predict 
the existence of dark matter, and for a resonable range of SUSY parameters, 
in an amount comparable to what is observed astronomically. 
These models have been waiting to be tested, and there are now a large 
number of experiments that are currently on line or will be on line in the 
relatively near future that will determine whether supersymmetry is a valid 
theory of nature.

While up to now the flow of information has been mostly from particle 
physics to cosmology, the new sattelite experiments \cite{2} MAP and Planck 
as well as about 25 balloon and ground based experiments can reverse this and 
can give constraints on what might be expected at accelerators. 
These astronomical experiments will measure the cosmological parameters 
with great accuracy. 
One uses the parameter  $\Omega_i=\rho_i/\rho_c$ where $\rho_i$ is the 
density of matter of type $i$ and $\rho_c= 3 H^2/8\pi G_N$ ($H=$Hubble 
constant, $G_N=$Newton's constant) to measure the mean amount of matter 
of type $i$ in the universe. 
Current measurements give $H= h\ (100\ km\ s^{-1} Mpc^{-1})$ with 
$0.5\alt h\alt 0.75$ (and hence $\rho_c= 1.88\times 10^{-29} h^2 gm/cm^3)$.
For a number of cosmological models one finds for cold dark matter (CDM) 
the range 
\begin{equation}
0.1 \le \Omega_{CDM} h^2 \le 0.4
\label{eq:1}
\end{equation}
and this range of CDM is the one assumed in many particle physics calculations. 
However, MAP and Planck measurements will greatly restrict this window, and 
impose strong constraints on the SUSY parameter space, and in this way restrict 
the SUSY mass spectrum. 
Thus astronomical measurements will impinge upon what is expected to be seen 
at accelerators. 

\section{Supergravity Models}
\label{sec:2}

In order to analyse dark matter predictions we need a well-defined SUSY model.
We will use here supergravity grand unified models with gravity mediated SUSY 
breaking (at a scale $\agt M_G$) in a hidden sector, and with gravity being the 
messenger of SUSY breaking to the physical sector \cite{3}. 
This gives rise to the following types of soft breaking masses at $M_G$ \cite{3,4}: 
$m_0$ (scalar soft breaking mass), $m_{1/2}$ (gaugino soft breaking mass), 
$A_0$ (cubic soft breaking parameter), $B_0$ (quadratic soft breaking parameter), 
as well as non-universal soft breaking terms \cite{5}.
In addition, a Higgs mixing parameter, $\mu_0$, for the two Higgs multiplets 
$H_1$ and $H_2$ ($W^{(2)}=\mu_0 H_1H_2$) can form naturally with 
$\mu_0\approx M_S$ \cite{5}.

In the following we will assume that the gaugino masses are universal at $M_G$ 
(as would be true to a good approximation for a simple GUT group). 
Over much of the parameter space $\mu^2\gg M_Z^2$ which leads to the ``scaling 
relations" for the charginos ($\chi_i^{\pm}, i=1,2$) and neutralinos 
($\chi_i^0, i=1-4$) \cite{6}:
$
2m_{\chi_1^0} \cong m_{\chi_1^{\pm}} \cong m_{\chi_2^0} 
\cong ({1\over 3} - {1\over 4}) m_{\tilde{g}}.
$

We assume here that the first two generations of squarks and sleptons are 
degenerate at $M_G$  (to suppress flavor changing  neutral currents) with 
common mass $m_0$ and parametrize the Higgs and third generation soft 
breaking masses at $M_G$ as:
\begin{equation}
m_{H_1}^2=m_0^2(1+\delta_1);\
m_{H_2}^2=m_0^2(1+\delta_2)
\label{eq:4}
\end{equation}
\begin{equation}
m_{q_L}^2=m_0^2(1+\delta_3); \
m_{u_R}^2=m_0^2(1+\delta_4);\
m_{e_R}^2=m_0^2(1+\delta_5)
\label{eq:5}
\end{equation}
\begin{equation}
m_{d_R}^2=m_0^2(1+\delta_6);  \
m_{l_L}^2=m_0^2(1+\delta_7)
\label{eq:6}
\end{equation}
Here $q_L\equiv (\tilde{u}_L, \tilde{d}_L)$  is the $L$ squark doublet, $u_R$ 
the $R$ up-squark singlet, etc. 
We limit our parameters to the domain 
$
m_0, m_{\tilde{g}} \le 1\ TeV; \ |A_t/m_0| \le 7; \
\tan\beta \le 25; \ |\delta_i| \le 1  
$
where $A_t$ is the $t$-quark $A$-parameter at the electroweak scale. 
These represent a choice of ``naturalness" conditions.
Note that for $\tan\beta \alt 25$, $\delta_5$, $\delta_6$, $\delta_7$ make 
only small contributions and can be neglected. 
(They would become important for larger $\tan\beta$.)

The RGE \cite{7} then give for $\mu^2$ at scale $M_Z$ the result \cite{8}:
\begin{eqnarray}
\mu^2=&&{t^2\over t^2-1}\biggl[\biggl\{ {1-3D_0\over 2}+{1\over t^2} \biggr\}+
   \biggl\{ {1-D_0\over 2}(\delta_3+\delta_4) - {1+D_0\over 2}\delta_2 
  + {1\over t^2}\delta_1 \biggr\} \biggr] m_0^2\nonumber\\
&&+ {t^2\over t^2-1}
\biggl[{1\over 2}(1-D_0){A_R^2\over D_0} + C_{\tilde{g}}m_{\tilde{g}}^2\biggr] 
- {1\over 2}M_Z^2\nonumber\\
&&+ {1\over 22}{t^2+1\over t^2-1}
\biggl(1+{\alpha_1\over \alpha_G}\biggr)S_0 + \ loop\ terms;\ \
t\equiv \tan\beta
\label{eq:8}
\end{eqnarray}
\noindent
Here $D_0\cong 1-(m_t/200\sin\beta)^2$, 
$A_R\cong A_t-0.613m_{\tilde{g}}$, $S_0= Tr Ym^2$ ($Y=$ hypercharge, 
$m^2=$ masses at $M_G$) and $C_{\tilde{g}}$ is given in Iba\~nez et al.\cite{7} 
$D_0$ vanishes at the $t$- quark Landau pole and hence for $m_t= 175\ GeV$ 
is generally small: $D_0\le 0.23$.
($A_R$ is the residue at the Landau pole.) 
Note that the choice $\delta_2<0$, $\delta_1>0$ and 
$\delta_3+\delta_4>0$ will increase the size of $\mu^2$, while 
$\delta_2>0$, $\delta_1<0$ and 
$\delta_3+\delta_4<0$ will decrease the size of $\mu^2$.
In the following we will see that $\mu^2$ plays a key role in dark matter 
predictions, and the effects of these particular sign possibilities of $\delta_i$ 
will allow a qualitative understanding of the phenomena.

\section{Direct Detection of Dark Matter}
\label{sec:3}

We review briefly in this section the direct detection of dark matter particles 
in the Milky Way incident on a terrestial detector. 
The density of such matter is estimated at $\rho_{DM}\cong 0.3\ GeV/cm^3$ 
with an impinging velocity of $v_{DM}\cong 300\ km/s$.
Calculation of detector event rates proceeds in two steps. 
One first calculates the expected $\chi_1^0$ relic density of 
$\Omega_{\chi_1^0}h^2$, left after annihilation in the early universe \cite{9}.
The size of $\Omega_{\chi_1^0} h^2$ varies as one moves across the SUSY 
parameter space.
In particular, for $m_{\chi_1^0}\alt 65\ GeV$ (or by the scaling relations, 
$m_{\tilde{g}}\alt 450\ GeV$) the annihilation cross section is dominated 
by the $s$-channel  $Z$ and $h$ poles and requires sensitive treatment 
\cite{10,11}.
For $m_{\chi_1^0}\agt 65\ GeV$, the $t$-channel  squark and slepton poles 
become dominant.
These two regimes will show up below in the detector event rates.

One first restricts the SUSY parameter space so that $\Omega_{\chi_1^0}h^2$ 
falls within the allowed window of the cosmological model under consideration. 
In addition, one restricts the parameter space so that accelerator bounds are 
obeyed. 
One then calculates the scattering of incident Milky Way $\chi_1^0$  by quarks 
in a nuclear target \cite{9} within the above restricted parameter space, 
giving rise to detector event rates $R$ measured in $ events/kg\ d$.

\section{Cosmological Models}
\label{sec:4}

Current astronomical measurements have sufficient uncertainty that they allow 
for a large variety of cosmological models. 
However, the Planck and MAP sattelites will be able to determine the basic 
cosmological parameters, $H$, $\Lambda$, $\Omega$ etc. to (1-10)\% accuracy 
by measurements of the deviations $\Delta T$ from the CMB temperature 
$T_0= (2.728\pm 0.002)\ {^{\circ}K}$.
Different cosmological models predict different angular correlations of the 
CMB deviations allowing for experimental 
determination of the cosmological model.
We consider here two possible cosmological models that might result from such 
measurements, and examine what consequences these might have on the sparticle 
spectra, and hence on accelerator searches for supersymmetry.

\noindent
{\it (i) $\nu$CDM Model}. If neutrinos have masses of $O(eV)$ they could furnish 
the hot dark matter (HDM) of the universe,  $\Omega_{\nu}$.
In addition there is baryonic dark matter ($B$) and CDM (the neutralinos $\chi_1^0$) 
such that the total $\Omega=1$.
As an example we assume that the sattelite measurements determine central 
values to be $\Omega_{\nu}=0.2$,  $\Omega_B=0.05$, $\Omega_{CDM}=0.75$
and a Hubble constant of $h=0.62$.
Using estimates of the errors with which these quantities can be measured by the 
Planck sattelite \cite{15,16} we find
\begin{equation}
\Omega_{CDM}h^2 = 0.288 \pm 0.013
\label{eq:14}
\end{equation} 
This shows the remarkable accuracy future determinations of cosmological 
parameters are capable of when compared with the current knowledge of 
Eq.\ (\ref{eq:1}). 

{\it (ii) $\Lambda$CDM Model}. As a second model we assume the existence of
a cosmological constant $\Lambda$ along with baryonic  and CDM, and assume 
that Planck has determined the central values of these parameters to be 
$\Omega_B=0.05$, $\Omega_{CDM}=0.40$, $\Omega_{\Lambda}=0.55$, and 
$h=0.62$.
Using the estimates of the errors in the Planck measurements \cite{15} one 
finds here that 
\begin{equation}
\Omega_{CDM}h^2 = 0.154 \pm 0.017
\label{eq:15}
\end{equation} 
While the above models are only two possible examples, they represent bounds 
on what might actually exist.
The important point is that the future balloon and sattelite measurements will be 
able to determine these parameters well, and the above models represent limits
within current knowledge.

\section{$\nu$CDM Model}
\label{sec:5}

  \begin{figure}
\begin{center}
\mbox{\psfig{figure=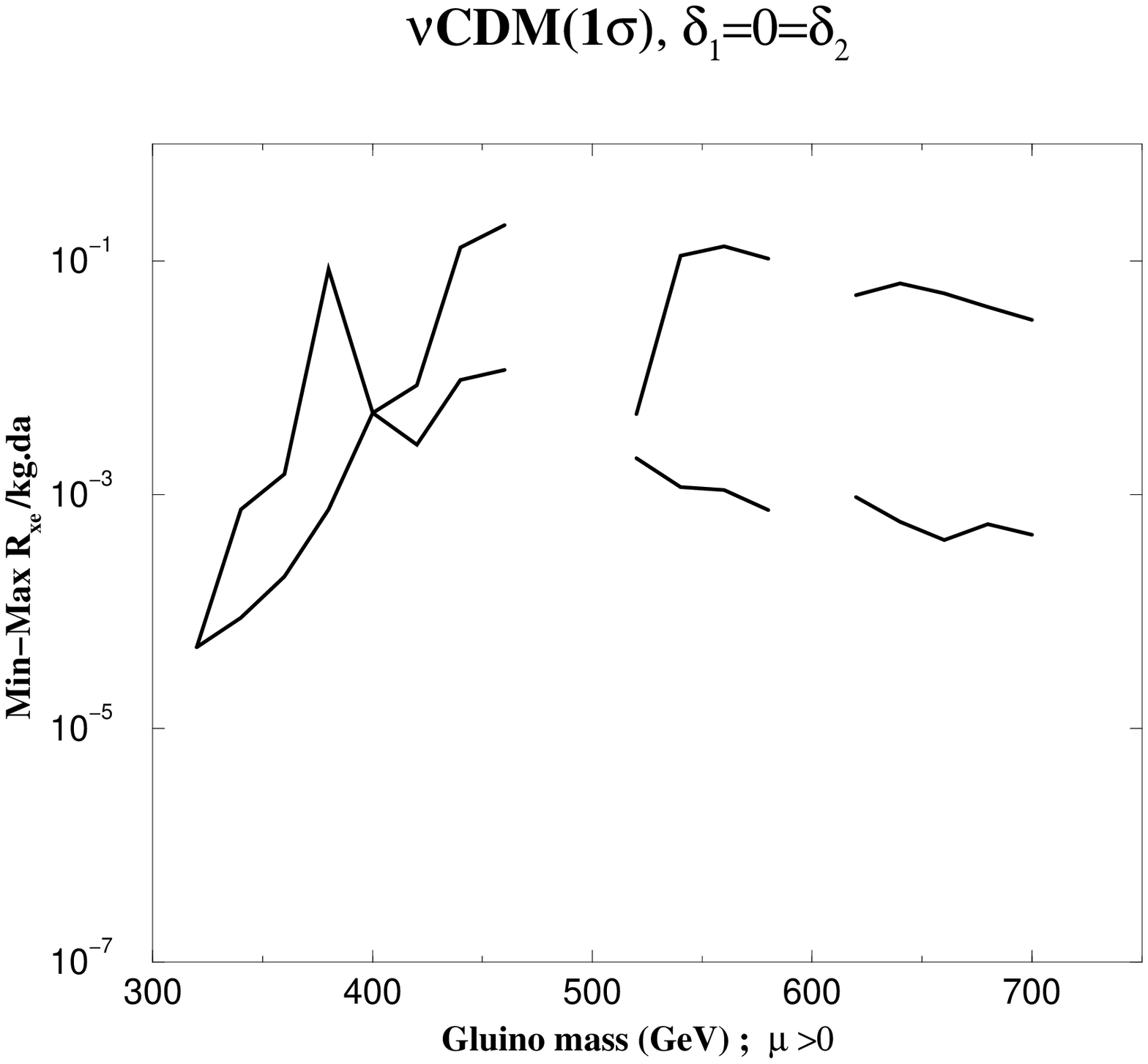,width=3in}}
\end{center}
 {\footnotesize 
Figure 1: 
Maximum and minimum event rates for a xenon CDM detector 
vs. gluino mass for $\nu$CDM model (1 std window) for universal 
soft breaking, $\mu>0$. From Ref.\cite{18}.
  } \end{figure}

We consider first universal soft breaking where by 
Eqs.\ (\ref{eq:4})-(\ref{eq:6}), the 
$\delta_i=0$. 
Fig. 1 plots the maximim and minimum event rates for Xe detector as a function of 
the gluino mass [$m_{\tilde{g}}\cong (7-8)m_{\chi_1^0}$ by the scaling relations] 
for a 1 std band of Eq.\ (\ref{eq:14}).
One may compare this with Fig. 2 of \cite{8}  where only the broad band of 
Eq.\ (\ref{eq:1}) is imposed.
One sees a reduction in event rates in the region $m_{\tilde{g}}\alt 450\ GeV$ 
(i.e.  $m_{\chi_1^0}\alt 65\ GeV$), the $Z$ and $h$ pole dominated region in the 
early  universe annihilation, and a corresponding increase for the higher masses. 
More significant is the appearance of forbidden 
regions, i.e. gaps, in the allowed values of $m_{\tilde{g}}$ in the region 
$m_{\tilde{g}}\cong 500\ GeV$ and $m_{\tilde{g}}\cong 600\ GeV$. 
In examining the effects of non-universal soft breaking, we note the correlation 
that when $\mu^2$ is decreased (which by Eq.\ (\ref{eq:8}) occurs for 
$\delta_2=1=-\delta_1$) the event rate $R$ is increased and when $\mu^2$ is 
increased ($\delta_2=-1=-\delta_1$) $R$ is decreased.
Fig. 2 shows these effects for the latter case, and also that the forbidden 
regions below $m_{\tilde{g}}=500\ GeV$ are significantly widened. 
For the alternate possibility ($\delta_2=1=-\delta_1$), the event rates increase. 
The maximum rates for this case are at the sensitivity of the current NaI detectors 
\cite{17},  and  detectors with a sensitivity of $R\agt 10^{-3}\ events/kg\ d$  
would cover the entire parameter space.
The gaps of the previous case have now disappeared. 
However,  $m_{\tilde{g}}$ is now bounded from below i.e. 
$m_{\tilde{g}}\agt 420\ GeV$.
In all of these cases there is also an upper bound on $m_{\tilde{g}}$ of 
$m_{\tilde{g}}\alt 720\ GeV$.
Thus cosmological constraints bound the allowed gluino mass range, and are 
sensitive to the non-universal soft breaking.

 \begin{figure}
\begin{center}
\mbox{\psfig{figure=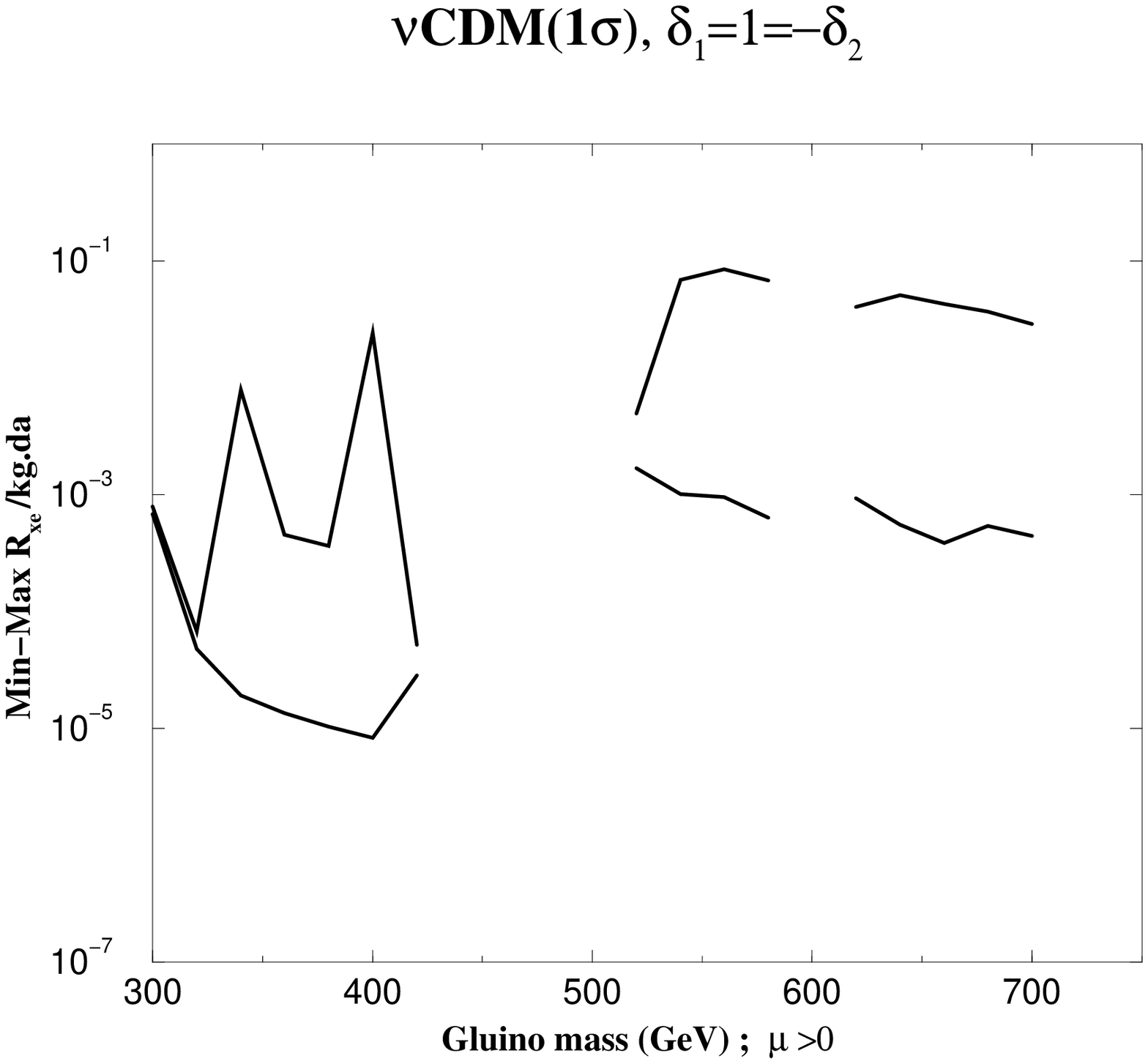,width=3in}}
\end{center}
{\footnotesize 
 Figure 2: 
 Same as Fig.1 for non-universal soft breaking with 
$\delta_1=1=-\delta_2$. From Ref.\cite{18}.
  } \end{figure}

The large $t$-quark mass causes large $L-R$ mixing in the stop $(mass)^2$ 
matrix, making the light eigenvalue, $m_{\tilde{t}_1}^2$, small. 
In general $m_{\tilde{t}_1}$ is governed by the size of $m_0$ and in the light 
neutralino domain $m_0$ can be large, i.e. $m_{\tilde{t}_1}$ can rise to 
$\approx 600\ GeV$. 
However, for heavy neutralinos ($m_{\tilde{Z}_1}\agt 65\ GeV$, 
$m_{\tilde{g}}\agt 450\ GeV$) the early neutralino annihilation is governed by 
the $t$-channel sfermion poles, and in order to get sufficient annihilation for
Eq.\ (\ref{eq:14}) to hold, $m_0$ must be small i.e. $m_0\alt 200\ GeV$. 
One finds in this domain that $m_{\tilde{t}_1}$ is generally an increasing 
function of $m_{\tilde{g}}$  with mass ranging from $250\ GeV$ to $500\ GeV$.

\section{$\Lambda$CDM Model}
\label{sec:6}

Here $\Omega_{\chi_1^0}h^2$ obeys Eq.\ (\ref{eq:15}) which even more sharply 
restricts the SUSY parameter space. 
Fig. 3 shows the maximum  and minimum expected event rates for a Xe detector 
for different values of $\delta_i$. 
While one has the same general behavior as in the $\nu$CDM model, there are 
interesting differences that distinguish the two.
Thus for this case for the 1 std window of Eq.\ (\ref{eq:15}), 
one finds a much tighter upper bound of $m_{\tilde{g}}\alt 520\ GeV$. 
There is also a lower bound  $m_{\tilde{g}}\agt 400 GeV$ for 
$\delta_2=1=-\delta_1$ (dashed curve), sharply restricting the gluino 
mass range for this case. 
Also, there are no gaps in the allowed gluino mass range. 

The squarks and sleptons are again constrained to be relatively light in the domain 
$m_{\tilde{g}}\agt 420\ GeV$ since here Eq.\ (\ref{eq:15}) requires 
$m_0\alt 100\ GeV$.
Thus $m_{\tilde{t}_1}$ ranges from $(200-400)\ GeV$ in this domain and for the 
first two generations of squarks, $m_{\tilde{q}}$ ranges from $400\ GeV$ to 
$500\ GeV$ and approximately scales with $m_{\tilde{g}}$. 
The lightest slepton, $\tilde{e}_R$ can have a mass  as low as $(85-90)\ GeV$  in 
this region, at the edge of detectability of LEP 190.

\begin{figure}
\begin{center}
\mbox{\psfig{figure=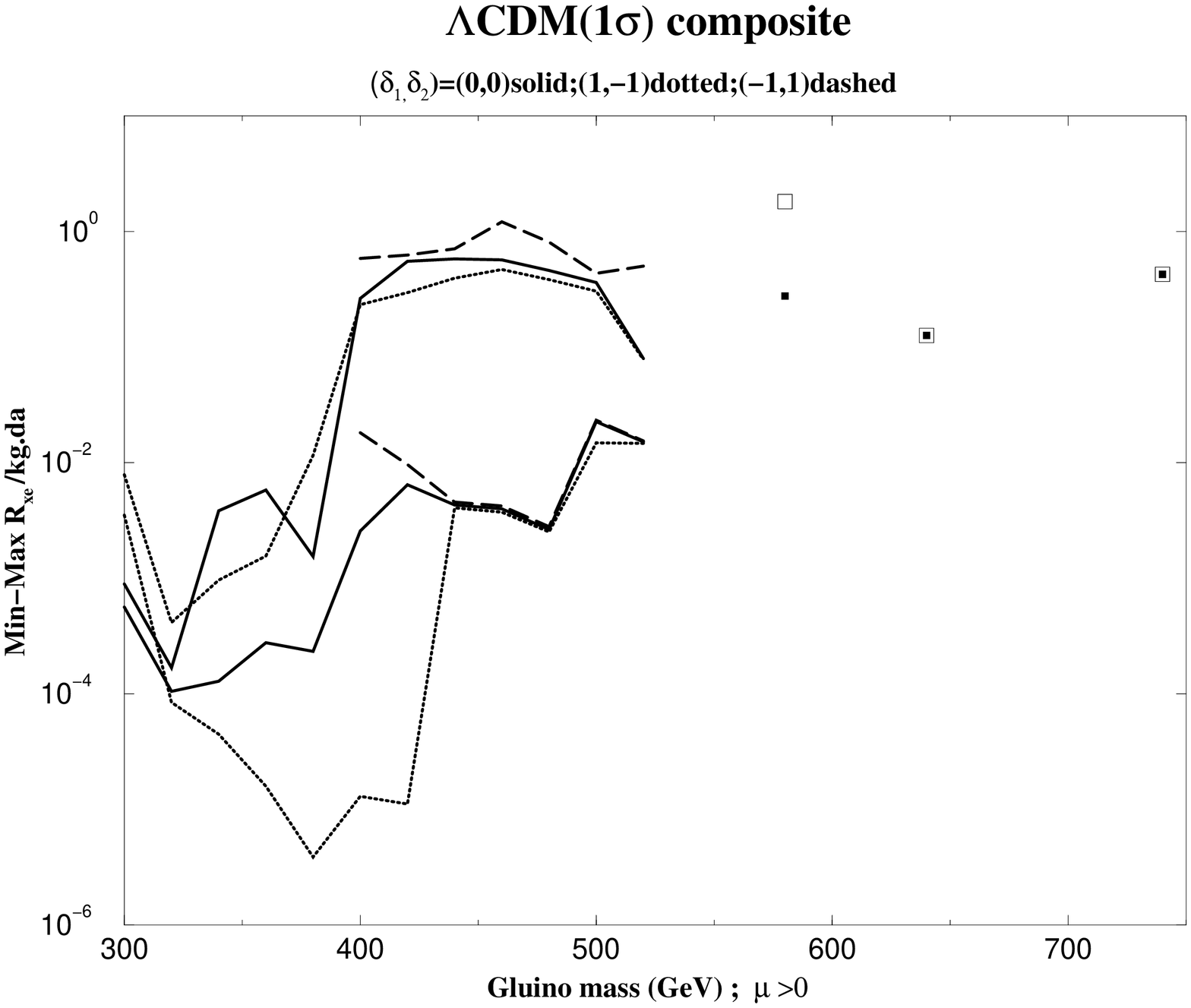,angle=270,width=3in}}
\end{center}
{\footnotesize 
Figure 3: 
Maximum and minimum event rates for a xenon CDM detector 
vs. gluino for $\Lambda$CDM model (1 std window), $\mu>0$, for universal 
soft breaking (solid curve) and non-universal soft breaking 
$\delta_1=1=-\delta_2$ (dotted curve), $\delta_1=-1=-\delta_2$ (dashed curve).
(The discrete high mass points are for $\delta_1=-1=-\delta_2$ where scaling is 
badly broken.) From Ref.\cite{18}.
} \end{figure}

This work was supported in part by National Science Foundation Grants 
 PHY-9722090 and PHY-96020274.

\end{document}